\documentclass[12pt]{article}

\usepackage[numbers]{natbib}
\usepackage{hyperref}       
\usepackage{times}
\usepackage{siunitx}
\usepackage{algorithm}
\usepackage{algorithmic}
\usepackage[version=4]{mhchem}
\usepackage[capitalise]{cleveref}
\usepackage{graphicx}
\usepackage[utf8]{inputenc} 
\usepackage[T1]{fontenc}    
\usepackage{booktabs}       
\usepackage{amsfonts}       
\usepackage{nicefrac}       
\usepackage{microtype}      

\newenvironment{sciabstract}{%
\begin{quote} \bf}
{\end{quote}}

\newcounter{lastnote}

\title{High-speed odour sensing using\\ miniaturised electronic nose}

\author
{
  \href{https://orcid.org/0000-0002-0380-2450}{\includegraphics[scale=0.09]{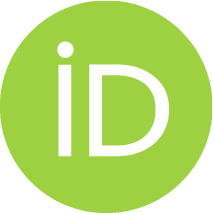} Nik Dennler},$^{1,2,\dagger}$
  \href{https://orcid.org/0000-0003-4107-5693}{\includegraphics[scale=0.09]{figures/orcid.eps} Damien Drix},$^{1}$
\href{https://orcid.org/0000-0002-2648-4264}{\includegraphics[scale=0.09]{figures/orcid.eps} Tom P. A. Warner},$^{3,4}$\\
  \href{https://orcid.org/0000-0002-2371-1935}{\includegraphics[scale=0.09]{figures/orcid.eps} Shavika Rastogi},$^{1,2}$
\href{https://orcid.org/0009-0004-9736-0121}{\includegraphics[scale=0.09]{figures/orcid.eps} Cecilia Della Casa},$^{3}$
\href{https://orcid.org/0000-0002-4964-1162}{\includegraphics[scale=0.09]{figures/orcid.eps} Tobias Ackels},$^{3,5}$\\
\href{https://orcid.org/0000-0002-4677-8788}{\includegraphics[scale=0.09]{figures/orcid.eps} Andreas T. Schaefer},$^{3,4}$
\href{https://orcid.org/0000-0001-6140-017X}{\includegraphics[scale=0.09]{figures/orcid.eps} André van Schaik},$^{2}$
\href{https://orcid.org/0000-0001-6753-4929}{\includegraphics[scale=0.09]{figures/orcid.eps} Michael Schmuker},$^{1,6,*}$\\
\\
\normalsize{$^{1}$Biocomputation Group, University of Hertfordshire, Hatfield AL10 9AB, United Kingdom}\\
\normalsize{$^{2}$International Centre for Neuromorphic Systems, Western Sydney University,}\\ \normalsize{Kingswood 2747 NSW, Australia}\\
\normalsize{$^{3}$Sensory Circuits and Neurotechnology Laboratory, Francis Crick Institute,}\\ \normalsize{London NW1 1AT, United Kingdom}\\
\normalsize{$^{4}$Departement of Neuroscience, Physiology and Pharmacology, University College London,}\\ \normalsize{London WC1E 6BT, United Kingdom}\\
\normalsize{$^{5}$Sensory Dynamics and Behaviour Lab, Institute of Experimental Epileptology and}\\ \normalsize{Cognition Research (IEECR), University of Bonn Medical Center, 53127 Bonn, Germany}\\
\normalsize{$^{6}$BioML Research Services, Berlin, Germany}\\
\\
\normalsize{\textsuperscript{\textdagger} E-mail:   dennler@proton.me}\\
\normalsize{$^\ast$ Corresponding author; E-mail:  m.schmuker@biomachinelearning.net}\\
}

\date{}

\begin{document} 


\baselineskip24pt


\maketitle 

\newpage
\begin{sciabstract}
Animals have evolved to rapidly detect and recognise brief and intermittent encounters with odour packages, exhibiting recognition capabilities within milliseconds. Artificial olfaction has faced challenges in achieving comparable results --- existing solutions are either slow; or bulky, expensive, and power-intensive --- limiting applicability in real-world scenarios for mobile robotics. 
Here we introduce a miniaturised high-speed electronic nose; characterised by high-bandwidth sensor readouts, tightly controlled sensing parameters and powerful algorithms. The system is evaluated on a high-fidelity odour delivery benchmark. We showcase successful classification of tens-of-millisecond odour pulses, and demonstrate temporal pattern encoding of stimuli switching with up to 60 Hz. 
Those timescales are unprecedented in miniaturised low-power settings, and demonstrably exceed the performance observed in mice. For the first time, it is possible to match the temporal resolution of animal olfaction in robotic systems. This will allow for addressing challenges in environmental and industrial monitoring, security, neuroscience, and beyond.  
 
\end{sciabstract}

\section*{Introduction}
The sense of olfaction is found all across the animal kingdom, and is crucial for survival and guiding behaviors such as navigation, food detection, predator avoidance, and mate selection \cite{brennan2004something, stensmyr2005insect, hallem2006insect, carde2008navigational, khan2012rats, sullivan2015olfactory, khallaf2021large}. Success in these tasks often hinges on the ability to swiftly and accurately detect and recognise scents \cite{riffell2008physical, van2014plume, van2015mosquitoes, demir2020walking}, particularly when dealing with odour plumes characterised by brief and intermittent encounters\cite{mafra1994fine, vickers2001odour} generated by turbulent dispersion processes \cite{Mylne1991, justus2002measurement, connor2018quantification}. 
Concentration fluctuations in odour plumes can exceed \SI{100}{\hertz} \cite{yee_vertical_1995}, while individual odour encounters can last single milliseconds or less \cite{celani_odor_2014} (see \cref{fig:enose}a). Many environmental cues are embedded in the fine structure of the odour plume \cite{hopfield_olfactory_1991, mafra-neto_fine-scale_1994, Schmuker_2016}, which various organisms have evolved to use for their advantage. 
For instance, \textit{Drosophila} olfactory receptor neurons can transduce odours in less than two milliseconds and resolve odour stimuli fluctuations at frequencies exceeding \SI{100}{\hertz} \cite{szyszka2014high}. 
Similarly, honeybee projection neurons decode odour identity in tens of milliseconds after stimulus onset \cite{krofczik_rapid_2008}, while mosquitoes can identify \ce{CO2} packets of just \SI{30}{\milli\second} \cite{dekker_moment--moment_2011}. 
A recent landmark study in mice has revealed their ability to discriminate rapid odour fluctuations, enabling them to distinguish temporally correlated from anti-correlated odours at up to \SI{40}{\hertz}, which facilitates source separation in complex environments \cite{Ackels2021}.

Research on mobile olfactory robotics \cite{s6111616} has flourished over the last decade; driven by promising applications and solutions across various domains \cite{francis2022gas}, and bootstrapping on the well established field of artificial olfaction\cite{Covington.2021}. The latter has demonstrated its effectiveness in domains where static and slow measurements are sufficient, such as the detection of hazardous gases or pollutants \cite{stetter1986detection}, spoilage alert systems \cite{maier2006monitoring}, health monitoring \cite{farraia2019electronic}, and food sciences \cite{jung_energy_2023}. However, many recent applications call for unmanned ground or aerial vehicles (UGV / UAV) to perform odour source localisation and navigation tasks \cite{kowadlo2008robot, burgues2019smelling, jing2021recent, francis2022gas}, which rely heavily on sensing the environment fast and efficiently, considering plume dynamics \cite{kadakia_odour_2022}.

Typically, mobile olfactory robots incorporate electronic noses; devices that are characterised by arrays of multiple gas sensors and associated peripheral electronics \cite{Persaud1982}. They offer distinct advantages over conventional analytical methods such as Photoionization Detectors (PID) and mass spectrometers (MS), notably in terms of portability, power efficiency, cost-effectiveness, and sensitivity to a wide range of odours and volatile compounds.
The most widely employed sensing components are Metal-oxide (MOx) gas sensors \cite{rock2008electronic}, which offer 
the significant advantage 
of a sensing layer that can be 
tuned through 1., modifications to its chemical structure, and 2., variations in operating temperature achieved by local heating, allowing for effectively detecting a diverse range of analyte classes.
Their minimal requirements for electronic peripheral components streamline sensor design, lower costs, and conserve valuable space. Further reductions in latency, form factor and power consumption were enabled through latest MEMS-based MOx sensors \cite{liu2018microhotplates, gardner_micromachined_2023}, facilitating seamless integration into electronic circuits \cite{ruffer2018new}.

However, the relatively slow response and recovery times of MOx sensor electronic noses pose challenges for widespread adoption, and are prohibitive for many potential robotic applications \cite{wang_odor_2022}. 
For this reason, various studies have investigated sensor response times and tried to improve them. Recent advancements in both hardware \cite{Vergara2014_hotplate, martinez2014using} and software \cite{monroy_overcoming_2012, di_lello_augmented_2014, Schmuker_2016, martinez_fast_2019, burgues_wind-independent_2019, burgues_high-bandwidth_2019, xing2019real, Drix2021} have significantly reduced response and recovery times from the orders of hours or minutes \cite{pashami_detecting_2012} down to tens-of-seconds or seconds \cite{burgues_high-bandwidth_2019, Drix2021}. 
Nevertheless, those timescales remain orders-of-magnitudes slower than what's observed for olfactory sensing in animals, potentially stalling progress on critical challenges in tracking of greenhouse gas emissions \cite{domenech-gil_electronic_2024}, ecological and environmental monitoring \cite{burgues_environmental_2020, tereshkov_metal_2024}, aerial-based wild fire detection \cite{Wang2023} disaster management \cite{fan_towards_2019}, and more.

In this work, 
we are pushing the limits of artificial olfaction with a high-speed, miniaturised electronic nose that can resolve odour pulses in the millisecond range. 
We propose an integrated e-nose design of MEMS-based MOx sensors and fast sampling periphery, as well as a set of powerful algorithms for control, sensing, and signal processing. 
We demonstrate the systems ability to operate at unprecedented temporal timescales when classifying short odour pulses, as well as when discriminating temporal characteristics of rapidly switching odour pairs. 
The challenge of deploying rapid and complex odour stimuli in a controlled and precise fashion \cite{marin_spatial_2021} is overcome by using a high temporal-precision olfactometer setup, which most recently has been used for showcasing the temporal odour recognition capabilities in mice \cite{Ackels2021, Dasgupta4278}. 

We first elaborate on the proposed design of the electronic nose and the feedback control methods, with which we achieve thermal response times that allow for ultra-fast heater cycles --- orders-of-magnitudes faster than what is suggested in the literature. 
Later, we show that the electronic nose can successfully classify the odour of short pulses, with durations down to tens of milliseconds. This is achieved by rapidly switching the sensor heater temperature, then extracting phase-locked data features to train machine learning classifiers. 
Further, we demonstrate the system's ability encode and infer temporal features in a task involving rapidly switching odour pairs, up to modulation frequencies of \SI{60}{\hertz}, which we show to match and even exceed the demonstrated capabilities of mice on equivalent tasks \cite{Ackels2021}. This is achieved by controlling the heater temperature to be constant, permitting for sensor response feature extraction from the frequency domain. 
Finally, we discuss our results and its implications, and identify some example use cases that may benefit highly from using fast sensing modalities. 

\section*{Results}
\subsection*{High-speed electronic nose and odour delivery system}
\begin{figure}[ht!]
    \centering
    \includegraphics[width=0.99\textwidth]{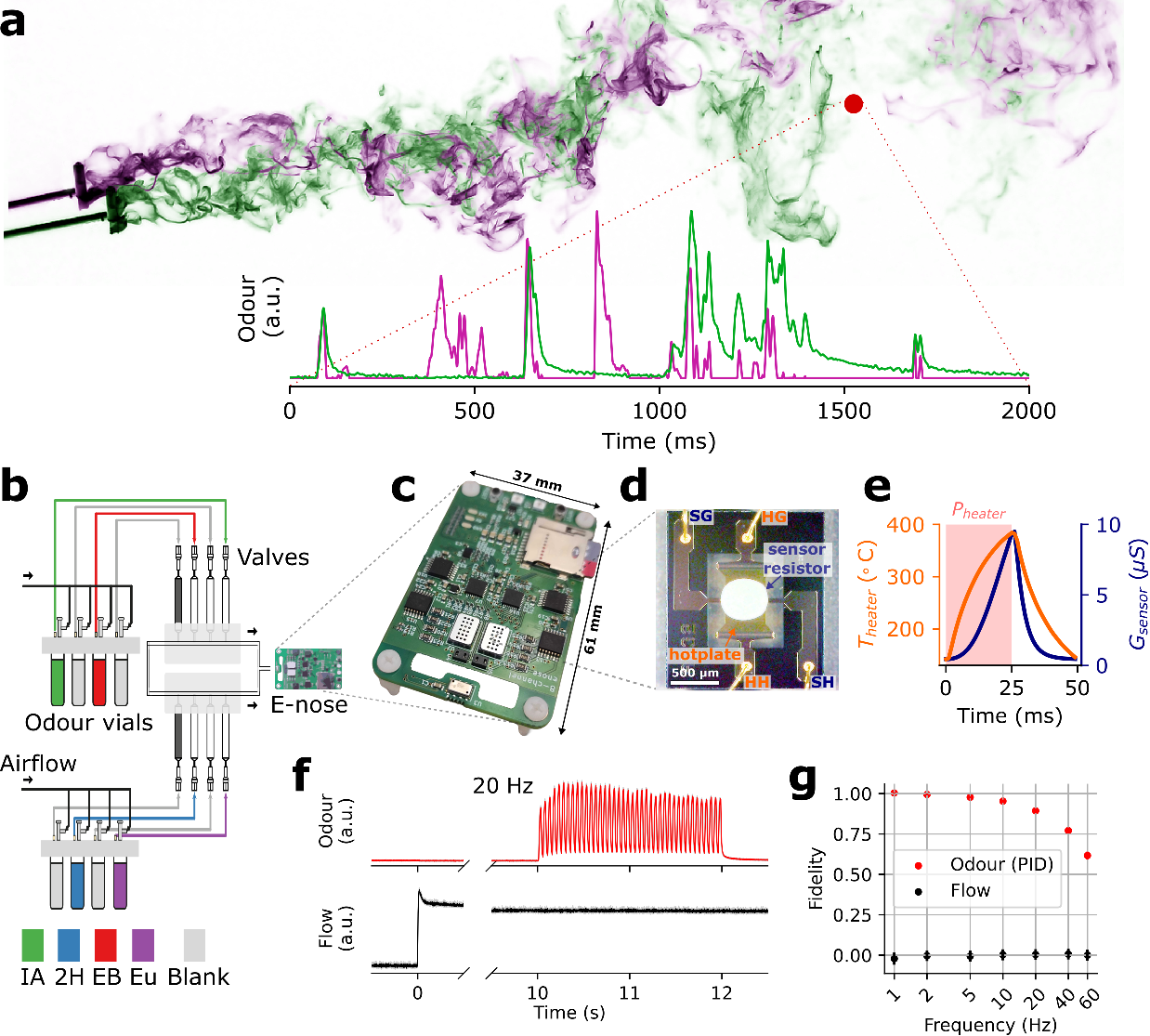}
    \caption{\textbf{Electronic nose and odour delivery system.} 
    \textbf{a}, Decoding temporal information of odour plumes requires fast sensing. Top: Two sequential \ce{TiCl4} smoke plume photographies, shifted and superimposed, kindly provided by Dr. Paul Szyszka. Bottom: Dual-PID recordings of source-separated odour plumes, from Ackels et al. \cite{Ackels2021}. Plume and sensor location (red) for illustrative purposes only. 
    \textbf{b}, Experimental setup with 
    odour delivery device and electronic nose. Adapted from Ackels et al. \cite{Ackels2021}.
    \textbf{c}, Electronic nose circuitry.  
    \textbf{d}, Microscopy image of the MiCS-6814 NH3 sensor with its housing removed. 
    \textbf{e}, Heater modulation cycle in ambient air. 
    \textbf{f}, PID and flow meter traces for a \SI{20}{\hertz} stimulus. Solid / faded (occluded) traces for mean / std. of five trials. 
    \textbf{g}, Resulting olfactometer temporal fidelity, for various frequencies.
    Odourants abbreviations: IA: isoamyl acetate; EB: ethyl butyrate; Eu: cineol; 2H: 2-heptanone; blank: odourless control.
    }
\label{fig:enose}
\end{figure}

We constructed a portable high-speed and high bandwidth electronic nose, which leverages the advantage of MEMS-based gas sensors and their rapid response times. We emphasised form factor and power consumption considerations that allow for sophisticated field measurements under space and power constraints, such as mobile robotic platforms\cite{jing2021recent}. 
Our design (\cref{fig:enose}c \& \cref{fig:supp-setup-methods}a) consisted of the following elements: a microcontroller for data processing and storage, eight analogue metal-oxide MEMS gas sensors (\cref{fig:enose}d), associated analogue circuitry and data converters, and a combined pressure, humidity and temperature sensor.

Ideal MOx sensor operation requires the sensing site to be heated to several hundred degrees. The sensor response is highly dependent on the temperature and its variation over time. Previous studies have shown that a modulation of the sensor's operating temperature often leads to better and faster gas discrimination performances \cite{Vergara2014_hotplate}, however the suggested sensor heater cycle durations were on the orders of seconds to minutes \cite{Vergara2014_hotplate, xing2019real, nakata2020characteristic, di2021optimizing, nakata2022distinction, Meng2023}. Aiming to achieve ultra-fast heater cycles, 

our design couples each sensor with a separate temperature control loop, which samples the temperature and adjusts the hotplate current at high frequency. 
\cref{fig:enose}d shows a typical heater modulation cycle in ambient air, where the sensor resistance follows the hotplate temperature in a low-pass fashion. In our experiments, we used two different heater temperature control schemes: one that cycled between low and high temperature values (\SI{150}{\degreeCelsius} and \SI{400}{\degreeCelsius}), 
and one at a constant high temperature (\SI{400}{\degreeCelsius}). 

To provide odour stimuli to the e-nose, we used an odour delivery system that can reliably present gaseous odour samples with a bandwidth beyond \SI{60}{\hertz}, described earlier \cite{Ackels2021, Dasgupta4278} and depicted in \cref{fig:enose}b. 
The system was based on high-speed micro-valves and incorporated a flow compensation mechanism \cite{Ackels2021}, ensuring exceptionally high temporal signal fidelity, and constant flow across the stimuli (\cref{fig:enose}f \& \cref{fig:enose}g and Methods).
As prototypical, simplistic high-frequency odour stimuli, we used square pulses of different duration and separation times. 
A set of odourants that resembles smells encountered in nature was considered: Ethyl butyrate (pineapple), isoamyl acetate (banana), cineol (eucalyptus) and 2-heptanone (cheese). The odourants were diluted in odourless mineral oil solvent. Additionally, we used two (identical) pure solvent samples as controls. The odours were presented as singular pulses with varying durations (\SI{10}{\milli\second} - \SI{1}{\second}) and concentrations (20\%-100\%), and as correlated and anti-correlated odour pulse trains (\SI{1}{\second}) at different modulation frequencies (2Hz - 60Hz). 

\subsection*{Rapid heater modulation enables robust data features}
\begin{figure}[t!]
    \centering
    \includegraphics[width=\linewidth]{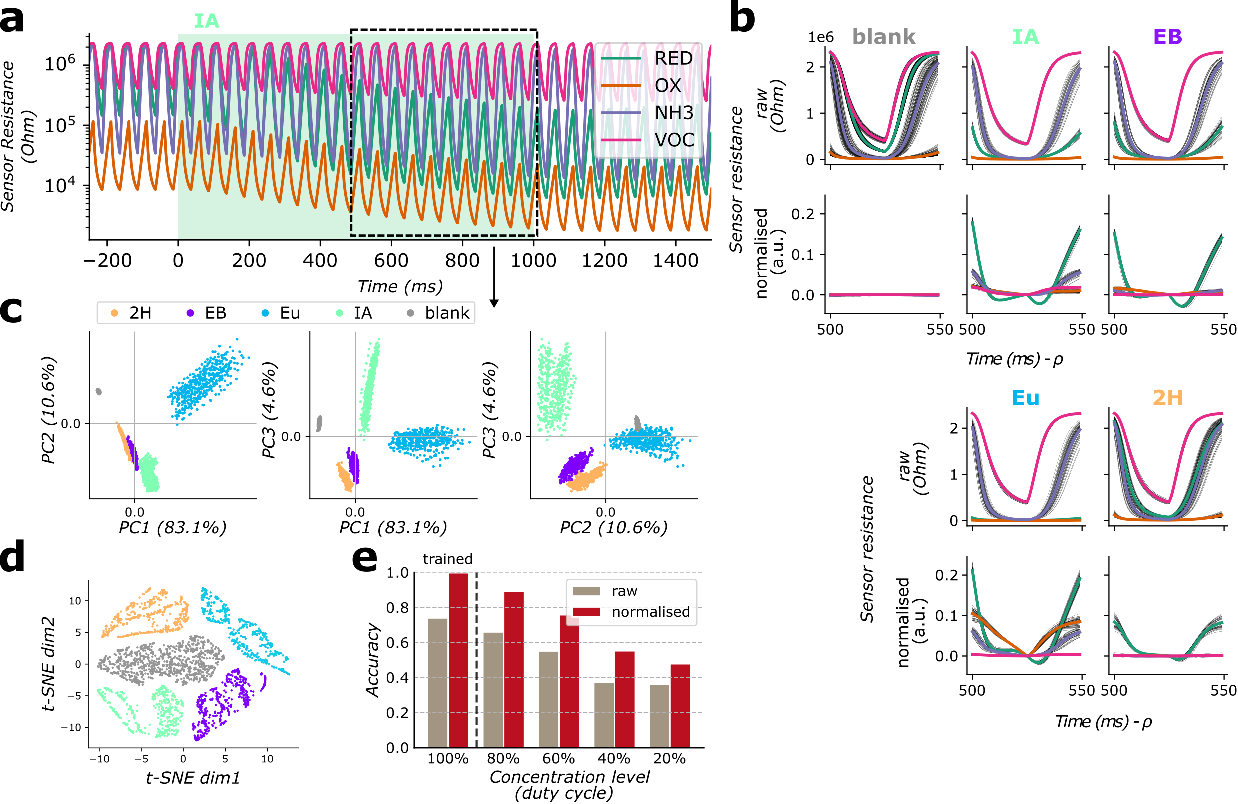}
    \caption{
    \textbf{Rapid heater modulation enables robust data features.}
    \textbf{a}, Sensor resistance of four MOx sensors with 20 Hz hotplate temperature modulation, responding to a \SI{1}{\second} odour pulse of isoamyl acetate (green background).
    \textbf{b}, \SI{50}{\milli\second} data feature for different gases, selected between \SI{500}{\milli\second} and \SI{550}{\milli\second} after odour pulse onset. Raw sensor response (upper) and normalised sensor response (lower, see Methods for normalisation procedure). Time shifted by cycle phase $\rho$ w.r.t. odour onset, for visual guidance only. 
    \textbf{c}, Principal component analysis (PCA), explained variance (most left) and projections, and 
    \textbf{d}, t-distributed stochastic neighbour embedding (t-SNE) visualisation, for the set of normalised data features extracted between \SI{500}{\milli\second} and \SI{1000}{\milli\second} after odour onset. 
    \textbf{e}, Accuracy scores for a k-nearest neighbours (k-NN) classifier trained on \SI{50}{\milli\second} data features from \SI{1000}{\milli\second} odour pulses at full concentration, and tested on \SI{50}{\milli\second} features from \SI{1000}{\milli\second} odour pulses at different concentration levels (tuned by adjusting the duty cycle of the micro-valves).
    }
    \label{fig:classfication_features}
\end{figure}

While cycling the sensor heater temperature can yield better odour classification results, the cycle duration may be restricting the temporal bandwidth at which a stimulus can be resolved.
In recent studies, we tested the effect of \SI{150}{\milli\second} duty cycles and found evidence for robust data features \cite{dennler2022_rapid, drixdennler2022_rapid}. In the current work, we leveraged our system's ability to rapidly modulate the sensor temperature, and cycled the heater temperature between a low step at \SI{150}{\degreeCelsius} and a high step at \SI{400}{\degreeCelsius} with a period of \SI{50}{\milli\second}. Notably, this is orders-of-magnitudes shorter than what had been suggested in previous studies \cite{Meng2023}. 
The resistance of the gas sensing elements closely tracks these changes in operating temperature (\cref{fig:enose}e \& \cref{fig:classfication_features}a), enabling us to extract gas features that are phase-locked with the heater cycles for subsequent analysis and classification. For this purpose, we divided the continuous stream of gas sensor samples into \SI{50}{\milli\second} chunks aligned with the temperature cycles (\cref{fig:classfication_features}b, upper row). The \SI{50}{\milli\second} data features further underwent pre-stimulus baseline normalisation and scaling (\cref{fig:classfication_features}b, lower; from now on referred to as "normalised data feature"); for details see Methods).

For testing class discriminability and robustness to concentration fluctuations, data features were extracted by sampling four sensors between \SI{500}{\milli\second} and \SI{1000}{\milli\second} after the onset of a \SI{1000}{\milli\second} odour stimulus. Principal Component projections (\cref{fig:classfication_features}c) and t-distributed stochastic neighbour embeddings (t-SNE) (and \cref{fig:classfication_features}d) show distinct clustering that coincided with odour classes. Further, a k-nearest neighbours (k-NN) classifier was trained on data features of one second long odour pulses at full concentration (100\%), and tested on features of one second long odour pulses at various concentrations (20\% - 100\%). 
The classification performance results are shown in \cref{fig:classfication_features}e. Notably, for the normalised data feature, 
the model provided $100\%$ classification accuracy at the trained concentration level, which remained at $(88.7 \pm 0.5)\%$ and $(81.2 \pm 0.6)\%$ when tested on $80\%$ and $60\%$ of the trained concentration level. Accuracy at lower concentration levels dropped significantly but remained well above chance. This is significantly better than what is achieved with the raw data feature (see \cref{fig:classfication_features}e). 

\subsection*{Time-resolved classification of millisecond odour pulses}
\begin{figure}[t!]
    \centering
    \includegraphics[width=\linewidth]{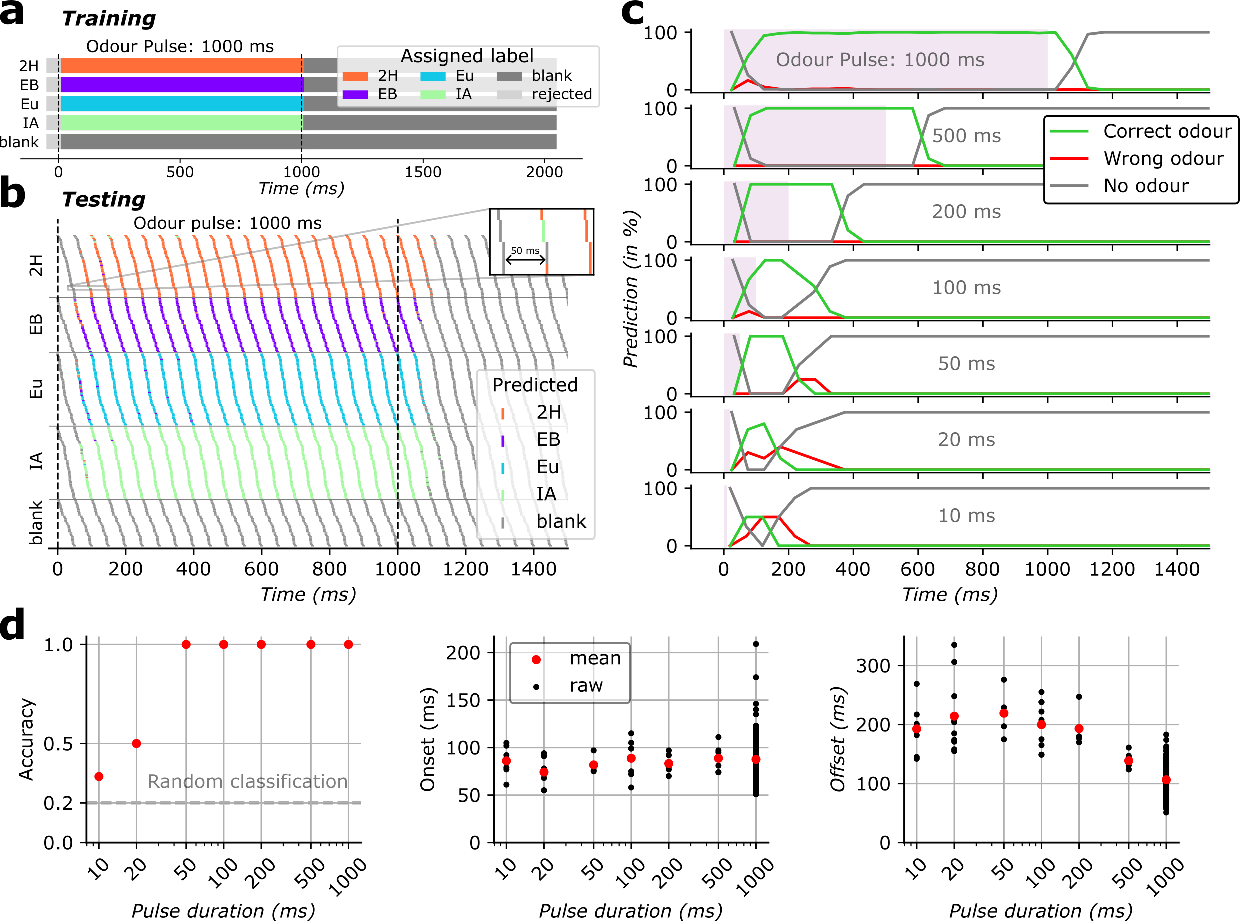}
    \caption{
    \textbf{Electronic nose can classify short odour pulses based on 50 ms data features.}
    \textbf{a}, Feature labels for the training set were phase aligned in relation to odour on- and offset.
    Features that overlapped with transition periods were not considered for training ("rejected", see Methods for parameters). 
    \textbf{b}, Odour stimulus classification over time for odour pulses of various lengths (\SI{10}{\milli\second} - \SI{1000}{\milli\second}, as predicted by a RBF-kernel SVM classifier trained on \SI{50}{\milli\second} features from \SI{1000}{\milli\second} second odour pulses. Shown here are \SI{1000}{\milli\second} pulses. 
    For visual clarity only, the trials are sorted by odour, and within each odour are sorted by phase w.r.t. stimulus onset. 
    \textbf{c}, Classification correctness over time (evaluated via the true odour presence), for different pulse durations. 
    \textbf{d}, Test accuracy, onset time and offset time for the prediction over time described in \textbf{b \& c}. Onset and offset were extracted using time-to-first non-'blank' and 'blank' prediction respectively, and shown here with respect to theoretical odour onset and offset.
    }
    \label{fig:fastclassification}
\end{figure}

In natural settings, odour bouts can be as brief as only milliseconds long. For an agent's successful interaction with the environment, this requires the ability to classify odours fast and robustly. 
We evaluated the ability of the electronic nose to classify odour pulses of various durations. A Support vector machine (SVM) with Gaussian radial basis function was trained on \SI{50}{\milli\second} data features, which were acquired from eight gas sensors throughout a \SI{1000}{\milli\second} odour stimulus at five concentration levels. Control trials ('blank') were included, obtained during a \SI{1000}{\milli\second} odourless mineral oil stimulation or immediately after odour pulses. See \cref{fig:fastclassification}a for a depiction of the labelled features. The trained model was deployed to predict the odour presence over time during exposure to odour stimulations of various durations, ranging from \SI{10}{\milli\second} to \SI{1000}{\milli\second}. \cref{fig:fastclassification}b displays the predicted classes over time on the example of an \SI{1000}{\milli\second} odour pulse, where \cref{fig:fastclassification}c summarises the predictions over time for all pulse durations. 

From these predictions, the corresponding accuracy, onset times, and offset times were derived and shown in \cref{fig:fastclassification}d. 
The classifier attained a 100\% accuracy in predicting the correct class for odour pulse durations ranging from \SI{1000}{\milli\second} down to \SI{50}{\milli\second}, despite not having been trained on pulses shorter than \SI{1000}{\milli\second}. Accuracy dropped for \SI{20}{\milli\second} and \SI{10}{\milli\second} pulses but remained above chance level. Notably, the classifier accurately and rapidly predicted the recovery of the sensor site, indicating 'no odour' when no odour was present. The time required for the classifier to correctly identify the odour remained relatively consistent across odour pulse durations, with an average value of (\SI{87}{\milli\second} ± \SI{20}{\milli\second}). Following odour offset, the classifier robustly predicted 'no odour' within (\SI{106}{\milli\second} ± \SI{24}{\milli\second}) for \SI{1000}{\milli\second} odour pulses. For shorter durations, this time increased inversely proportional to the pulse duration, which can presumably be attributed to the sensor's integration time that may approach or exceed the duration of the short odour pulses.

\subsection*{Decoding temporal structure of rapidly switching odours}
\begin{figure}[t!]
    \centering
    \includegraphics[width=0.99\linewidth]{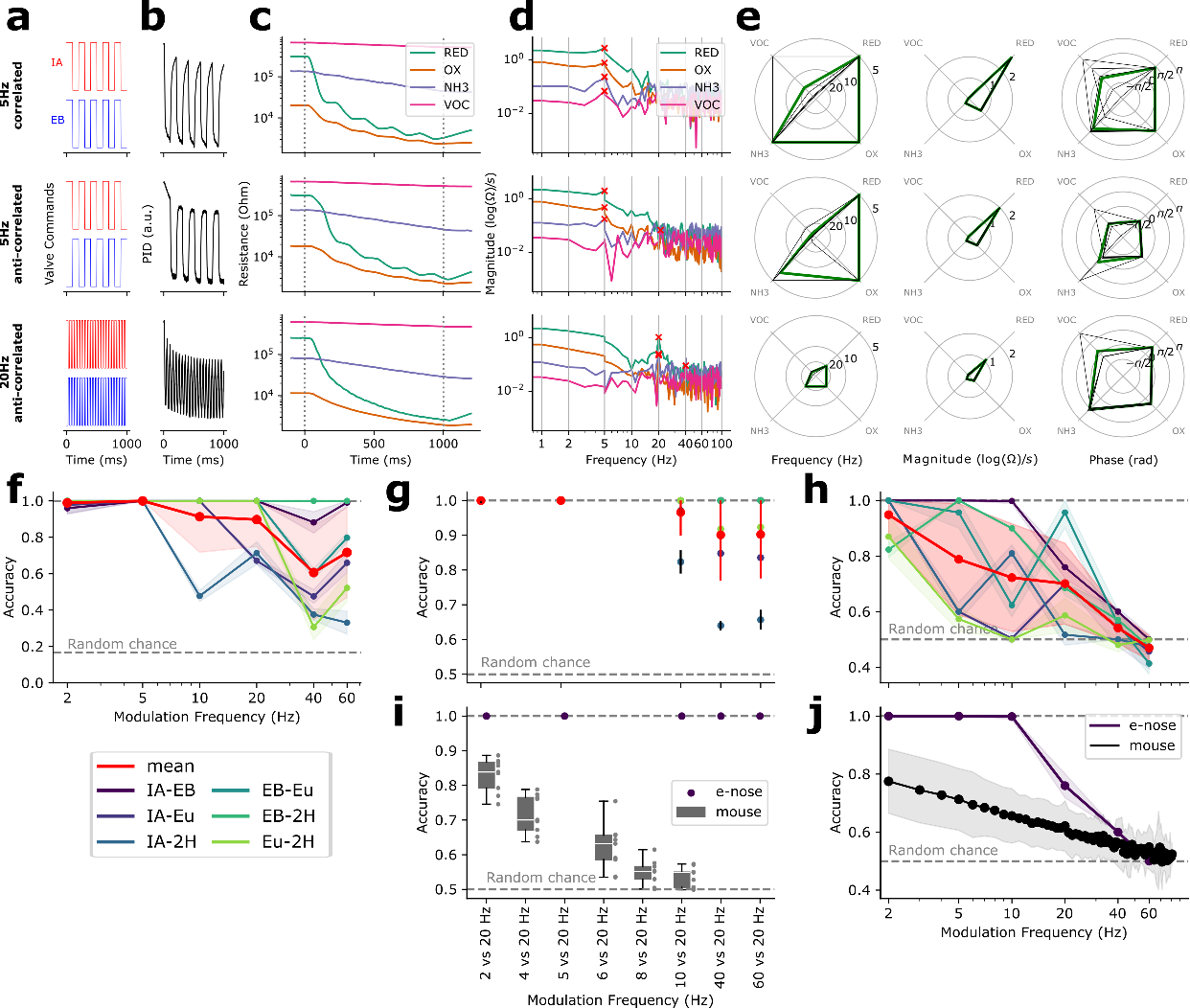}
    \caption{
    \textbf{Decoding temporal structure of rapidly switching odours}.
    \textbf{a,} Odour valve commands.
    \textbf{b,} PID response. 
    \textbf{c,} Electronic nose response. 
    \textbf{d,} Discrete Fourier Transformation (DFT) of first derivative of the sensor log-resistance. Crosses denote highest-magnitude peaks. DFT bin frequencies were rounded to nearest integers, for visual clarity only. 
    \textbf{e,} Feature visualisation frequency, magnitude and phase of the dominant DFT peaks. Thick lines; means of corresponding trials, thinner lines; single trials.
    \textbf{f}, Class-balanced accuracies for modulation frequency classification. 
    \textbf{g}, Accuracies for binary modulation frequency classification. 
    \textbf{h}, Class-balanced accuracies for binary modulation mode classification (corr. vs. anti-corr.).
    \textbf{i}, Subset of \textit{g} for IA-EB, for mouse performance comparison (described in detail in Ackels et al.\cite{Ackels2021}). 
    \textbf{j}, Subset of \textit{h} for IA-EB, for mouse performance comparison. 
    Panels \textit{a}-\textit{e} show representative trials only.
    For \textit{f}-\textit{j}, electronic nose accuracy mean and SD (clipped at 1.0) arise from repeated training and testing with different random seeds. 
    }
    \label{fig:temporal_combined}
\end{figure}
In the presence of multiple odours, detecting whether the odour encounters are correlated or not can help to infer whether they come from the same source or from separate locations \cite{hopfield_olfactory_1991}. Further, information about the encounter frequency can give rise to spatial source information \cite{Schmuker_2016}.
It has been shown that mice can distinguish between correlated and anti-correlated odour pairs reliably up to correlation frequencies of \SI{40}{\hertz} \cite{Ackels2021}; a feat that has not yet been matched in robotic systems. Considering performance metrics based on similar tasks, here we explored the ability of the electronic nose to resolve temporal structure of odour stimuli.
Rapidly alternating odour pairs were presented at frequencies between \SI{2}{\hertz} and \SI{60}{\hertz} for a duration of one second. 
We discriminated between two odour pulse trains being either in phase (correlated) or shifted by half a cycle (anti-correlated) (\cref{fig:temporal_combined}a). 
The resulting odour patterns follow the pulses rapidly (PID recordings in \cref{fig:temporal_combined}b). 

While heater modulations lend themselves for extracting phase-locked data features and thus allowing for efficient odour classification, maintaining the sensor heater temperature constant instead allows for analysing the data in continuous time. Particularly when observing repeating patterns or complex temporal dynamics of a stimulus this may be advantageous, as the sensor response can be regarded in its frequency domain. 
Thus, for the following experiments, we operated four MOx sensors of the electronic nose under a constant hotplate temperature of \SI{400}{\degreeCelsius}. The sensors responded to the stimuli by dropping their resistance at the pulse-train onset, with the stimulus modulation visually embedded in the response (\cref{fig:temporal_combined}c). We extracted data features by differentiating and logarithmically scaling the raw sensor response, followed by a discrete Fourier transform (DFT) (\cref{fig:temporal_combined}d). For each sensor, where the maximal magnitude is found, the frequency, the magnitude and the phase were extracted. This yielded a 12-dimensional data feature (\cref{fig:temporal_combined}e). A visual comparison of the features reveals distinct differences between correlated and anti-correlated pulse trains (top vs. middle), as well as between different frequencies (middle vs. bottom).

On those features, ensembles of Random Forest classifiers were trained for three tasks: 1. Decoding the modulation frequency of two odour pulse trains from a set of frequencies, 2. predicting the modulation frequency of two odour pulse trains from pairs of frequencies, and 3. decoding if two odours pulse trains are either correlated or anti-correlated.  
For the latter two tasks and a subset of the odours, a comparison with the mouse performance as detailed in Ackels et al. \cite{Ackels2021} is provided. 

The test accuracies for the three tasks and all gas combinations are shown in \cref{fig:temporal_combined}f-h. For task 1, the data recorded with the electronic nose enables nearly perfect frequency classification (\cref{fig:temporal_combined}f) for modulation frequencies up to \SI{5}{\hertz}, then on average decreasing to $(0.91 \pm 0.20)$  and $(0.90 \pm 0.15)$ for  \SI{10}{\hertz} and \SI{20}{\hertz} respectively, and finally dropping to $(0.61 \pm 0.26)$  and $(0.72 \pm 0.25)$ for  \SI{40}{\hertz} and \SI{60}{\hertz} respectively. 
For the pair-wise frequency classification (task 2, \cref{fig:temporal_combined}g), classification performance is perfect for modulation frequency pairs \SI{2}{\hertz} vs \SI{20}{\hertz} and \SI{4}{\hertz} vs \SI{20}{\hertz}, decreasing to accuracies of 
$(0.97 \pm 0.07)$, $(0.90 \pm 0.13)$ and $(0.90 \pm 0.13)$
for the pairs \SI{10}{\hertz} vs \SI{20}{\hertz}, \SI{40}{\hertz} vs \SI{20}{\hertz} and \SI{60}{\hertz} vs \SI{20}{\hertz}, respectively. 
For discriminating correlated vs. anti-correlated pulse trains (task 3, \cref{fig:temporal_combined}h), it appears that the electronic nose, on average, enables high prediction scores of $(0.95 \pm 0.08)$, $(0.79 \pm 0.20)$, $(0.72 \pm 0.19)$ and $(0.70 \pm 0.14)$ for modulation frequencies of \SI{2}{\hertz}, \SI{5}{\hertz}, \SI{10}{\hertz} and \SI{20}{\hertz}, respectively. This drops to $(0.54 \pm 0.06)$ for \SI{40}{\hertz} and finally to $(0.47 \pm 0.05)$ for \SI{60}{\hertz}.

In Ackels et al. \cite{Ackels2021}, the odour pair isoamyl acetate --- ethyl butyrate (IA-EB) has been used to test the discrimination power of fast odour dynamics in mice. In the following, we consider the corresponding subset of the electronic nose recordings and compare them to the named study.
For the pair-wise frequency classification (task 2, \cref{fig:temporal_combined}i), the electronic nose classification performance is perfect for all the tested modulation frequency pairs, from \SI{2}{\hertz} vs \SI{20}{\hertz} up to \SI{60}{\hertz} vs \SI{20}{\hertz}. Here, the mouse performed significantly worse --- for the pair \SI{2}{\hertz} vs \SI{20}{\hertz}, the mouse accuracy score was $(0.83 \pm 0.05)$ and then progressively dropped down to  $(0.53 \pm 0.03)$ for \SI{10}{\hertz} vs \SI{20}{\hertz}.
Finally, considering the results from the phase prediction task (task 3, \cref{fig:temporal_combined}j), it appears that the electronic nose enables perfect prediction scores up to modulation frequencies of \SI{10}{\hertz}, which then steeply drops to an accuracy of $(0.76 \pm 0.05)$ for \SI{20}{\hertz}, $(0.60 \pm 0.00)$ for \SI{40}{\hertz}, and finally to chance level for \SI{60}{\hertz}. In comparison, the mouse scores $(0.78 \pm 0.11)$ at a modulation frequency of \SI{2}{\hertz}, linearly decaying in accuracy down to chance level at around \SI{80}{\hertz}. 

In order to validate if the observed performance can be attributed to the odour signal, and not to potential artefacts caused by potential hotplate temperature variations (which may be caused by unnoticed flow fluctuations), we repeated the experiment using the hotplate temperature signal (see \cref{fig:supp-temporaldiscrimination-methods}i), as well as the PID responses (see \cref{fig:supp-temporaldiscrimination-methods}m). In both cases, the analogous feature extraction and classification pipeline was performed, resulting in classification performances as displayed in \cref{fig:supp-temporaldiscrimination-methods}j-l for the hotplate temperature, and \cref{fig:supp-temporaldiscrimination-methods}n-p for the PID responses. The analysis confirmed that there was not enough information in the hotplate temperature response alone to classify the odourants with above chance performance. Further, using the PID response, which should be unaffected by potential flow fluctuations and is commonly used as a ground-truth measurement, nearly perfect accuracy scores were achieved for most gas combinations across the tested tasks. 

\section*{Discussion}
For many tasks and applications in robotics, natural and turbulent environments pose the challenge of highly dynamic and rapidly changing odour concentrations, which demands high temporal resolution odour sampling and processing. 
Intrigued by the exceptional speed at which animals process and respond to odours, 
we challenged the limits of artificial olfaction by introducing and evaluating a portable and low-power high-speed electronic nose. For this, we coupled an integrated design of MEMS-based MOx sensors and fast sampling periphery with a set of highly optimised algorithms for control, sensing, and signal processing. 
To assess its capabilities, the electronic nose was subjected to odour stimuli delivered through a high-bandwidth system. Odours were presented in the form of square pulses with varying durations and concentrations, or as pairs of pulse trains at varying frequencies and phase. 

We demonstrated that the electronic nose can successfully infer the odour identity of single-odour pulses down to durations of \SI{10}{\milli\second}, albeit being trained on \SI{1}{\second} odour pulses only. This was achieved through modulating the sensor temperature with cycle periods of \SI{50}{\milli\second}, extracting and pre-processing the corresponding sensor response, then training and evaluating a classifier.
Further, we demonstrate the electronic nose's ability to predict whether two-odour pulse trains were correlated or anti-correlated up to switching frequencies of \SI{40}{\hertz}; matching or exceeding mice on the equivalent task. For tasks involving determining the odour switching frequency (multi-class and binary), we demonstrate a high performance up to \SI{60}{\hertz}, outperforming mice on equivalent tasks. For this, the sensor heaters were set to provide a constant temperature, which allowed an analysis of the data in the frequency domain. 

Some valuable insights on task-specific sensing modes and sampling can be gained from those results. First, data windows that are phase-locked with ultra-short sensor temperature cycles appear to be a sensible choice when given the task of odour classification. The high sampling rate allows for a multi-dimensional data feature (here, 8 sensors times 50 samples per feature), which --- together with the pre-stimulus normalisation procedure --- successfully captures the odour-specific sensor response. In mammalian sensory neuroscience, the analogy would be the phase-coupling of spike-trains to the inhalation cycle, allowing the spike-timings to encode information about the odour identity, thus suggesting one "sniff" as the unit of olfactory processing \cite{kepecs_sniff_2006}. 
Conversely, if the task is not classification but decoding temporal information about the odour stimuli, such as frequency or correlations, integrating the information across an artificial time window would limit the temporal resolution, hence recording continuously and without heat modulation might be the better choice. The ability of mammals to access temporal stimulus information at sub-sniff resolution has been demonstrated \cite{Ackels2021, Dasgupta4278}, and shown to be relevant for behavioural tasks \cite{Ackels2021}. 
Those findings may suggest future experiments in which both modes are active simultaneously on separate sensor instances --- continuously-sampled constant-temperature and time-integrated temperature-modulated --- which could allow for extracting information about the temporal profile and the identity in parallel. Research on insect have suggested dual-pathway olfactory systems \cite{galizia_parallel_2010}, which may facilitate the simultaneous extraction of odour identity and concentration information \cite{schmuker_parallel_2011}. Such an approach might suggest elegant solutions to the olfactory cocktail party problem \cite{rokni_olfactory_2014}.

The proposed technology and its evaluation 
hold promise for tackling many real-world challenges that require rapid odour sensing. 
In particular, any instance of olfactory robotic solutions might currently be cut short in terms of performance; as for both UGVs and UAVs, the sensor response time dictates the maximum speed at which the agent can move while still obtaining spatially resolved measurements \cite{burgues_environmental_2020}. Thus, such applications may directly benefit from using the proposed sensor modalities, allowing for faster identification and localisation of odour sources. 
For instance, a recent work proposed swarms of nano quadcopters performing gas source localisation in indoor environments \cite{duisterhof_sniffy_2021}, and evaluated different search strategies. A decreased latency in detection and classification may not only assist in more efficient source localisation, but also in expanding the use case to multiple odours and more complex outdoor environments. 
Another recent study proposes odour sensing on drones for wildfire monitoring \cite{Wang2023}. For detecting smoke; vision and gas sensors are fused, however the gas sensor update frequency is just \SI{1}{\hertz}. Given the intermittent and fine-structured nature of odour plumes, an improvement on sensing timescales could reduce false-negatives and aid in gaining critical time in localising the fire. 

Beyond robotics, most applications in security still use static and relatively slow sensing platforms, for e.g. the odour-based detection of explosives \cite{9133565} at airports. At checkpoints, fast and portable electronic noses could replace random spot checks with exhaustive controls, and thus minimise risk further.
Further, recent investigations on mammalian olfactory-guided behaviour use head-mounted MOx sensors as control recordings \cite{tariq_using_2021}. Using a high-resolution data acquisition system --- particularly one that matches the temporal capabilities of the subject --- would allow for better data quality and hence could improve the resulting models. 

In a related vein, neuromorphic information processing \cite{mead_neuromorphic_1990, indiveri_neuromorphic_2011} has seen much traction in recent years, where in particular the reduced latency, power consumption and data bandwidth have enabled highly optimised vision and auditory sensors \cite{lichtsteiner2008128, liu2010neuromorphic}. We suggest that the information embedded in millisecond odour packets, together with the sparse and intermittent nature of odour plume encounters, makes the sense of olfaction an ideal candidate for neuromorphic sensing. We foresee that revealing the rapid nature of the sensors will further stimulate this field of research, motivating event-driven and asynchronous odour sampling \cite{persaud2013neuromorphic}, for MOx sensors \cite{chicca2014neuromorphic, rastogi2023spike} and beyond \cite{laplatine_silicon_2022, wang_biomimetic_2024}.

In conclusion, our study marks a groundbreaking advancement in electronic olfaction systems, demonstrating the ability to discern odours and decode odour patterns with unprecedented temporal precision in miniaturised low-power settings. 
Our findings unlock new possibilities for developing robotic systems capable of rapidly and precisely tracking odour plumes in compact and low-power environments, with the potential to transform electronic nose designs and their applications across various domains.

\subsection*{Data and code availability}
The dataset and analysis code will be made publicly available upon publication, but can be provided earlier on request. 

\section*{Methods}
\label{sec:methods}

\subsection*{Electronic nose}

\subsubsection*{Circuit board design}
The electronic nose uses readily available components and is illustrated in Figure \cref{fig:supp-setup-methods}a. It features a \textit{Raspberry Pi Pico} microcontroller and incorporates eight MEMS-fabricated MOx gas sensors of four different types, grouped into four packages. 

The sensor packages comprise two \textit{SGX Sensortech} MiCs-6814 sensors (sensors 1 and 5) 
and two \textit{ScioSense} CCS801 sensors (sensors 2,3,4 and 6,7,8) 
, capable of detecting a wide range of reducing and oxidising gases, including volatile odour compounds (VOCs), hydrocarbons, carbon monoxide, hydrogen, nitrogen oxides, and ammonia. Figure \cref{fig:enose}c displays an optical microscopy image of the sensor structure. For sensing, the DC resistance across the sensing element is measured. The integrated micro-hotplates allow for operation at temperatures of up to \SI{500}{\degreeCelsius}. 

For controlling the micro-hotplates, eight operational amplifiers (2x \textit{STMicroelectronics} TS924) 
are employed, and the sensor heater voltage is configured using two Digital-to-Analogue Converters (DACs), specifically the \textit{Texas Instrument} DAC60004
, which offer four channels each at 12 bits and 1 kHz. Additionally, two Analogue-to-Digital Converters (ADCs), the \textit{Texas Instrument} ADS131M08
, are used to read sensor and heater resistances. These are differential, simultaneous-sampling ADCs which read out the eight gas channels and the eight temperature channels in lockstep at 24 bits and 1 kHz (the two ADCs share the same clock). To monitor environmental conditions, a digital pressure-humidity-temperature sensor, the \textit{TE Connectivity} MS8607
, is included, which samples data at 24/16/24 bits and 50 Hz. Real-time data logging is facilitated through the inclusion of a microSD card.
The device's power needs, ranging from \SI{1.2}{\watt} to \SI{1.5}{\watt}, allow for multiple days of continuous operation on a pocket-sized battery pack, making it suitable for extended field recordings or robotic environments.

\subsubsection*{Sensor heater modulation}
To implement controlled heater modulation, continuous measurement of the hotplate temperature and regulation of power delivered to the resistive heating element are essential. Each heater voltage \(V_{\text{heat}}\) was adjusted using a DAC and an associated amplifier, while the resulting current \(I_{\text{heat}}\) was monitored using an ADC in conjunction with a fixed-value sense resistor \(R_{\text{sense}}\). From these two quantities one can compute the heater resistance \(R_{\text{heat}} = V_{\text{heat}}/I_{\text{heat}}\) and dissipated power \(P_{\text{heat}} = V_{\text{heat}}I_{\text{heat}}\). Because the device did not directly measure \(V_{\text{heat}}\), the resistance calculated by substituting the known control and sense voltages \(V_{\text{dac}} - V_{\text{sense}} \approx V_{\text{heat}}\) was subject to errors, of which transient errors caused by lag and settling time in the DAC and amplifier were deemed the most significant. These affected the sample acquired immediately after a change in control voltage; therefore we used a Kalman filter \cite{welch_introduction_1995} to estimate \(R_{\text{heat}}\), setting the measurement uncertainty proportionally to the rate of change of the control voltage \(V_{\text{dac}}\).

The kind of resistive heating element present in our design exhibits a quasi-linear relationship between hotplate resistance and temperature \cite{rastrello2010thermal}, so we used a linear model to map the heater resistance to a calculated hotplate temperature. The parameters of that linear model were set before recording the data by measuring the heater response to a series of power steps, and matching it with calibration data from the manufacturer's datasheet, namely the nominal hotplate temperature delta above ambient air temperature at nominal heating power. This calibration procedure was repeated on an approximately weekly basis to account for the possible ageing of the sensors.

Achieving short temperature steps of a duration not much greater than the thermal time constant of the hotplate presented a number of challenges. Because the shortest steps consisted of only 25 samples (\SI{25}{\milli\second} at \SI{1}{\kilo\hertz}), we employed a combination of open-loop and closed-loop control. In that scheme, we operated the heater at a constant voltage for the duration of each step. That voltage was selected based on a learned mapping between a desired temperature change, and the control voltage required to achieve it. We adjusted the mapping after every step to compensate for the effects of airflow and ambient temperature fluctuations, using a proportional controller with a relatively slow adaptation rate (\SI{0.1}{\volt\per\degreeCelsius\per\second}). With this, fast and repeatable temperature modulation patterns can be obtained without introducing artefacts due to e.g. control loop oscillations.

For experiments with a constant heater temperature, achieving fast temperature changes was not an issue, but care was taken to avoid the artefacts caused by the DAC's 12-bit quantisation of applied heater voltage. These quantisation steps of about \SI{0.7}{\milli\volt} led to small but measurable transients in the recorded sensor signal. Since these transients were in the same frequency band as the signals of interest, we decided to also keep the heater voltage constant during each stimulus. We adjusted it to eliminate the temperature error after each stimulus, which was sufficient since the thermal environment of the sensors changed only slowly.

\subsubsection*{Sensor responses to ambient air}
The resistance of MOx sensors depends not only on the presence of gases, but also on the operating temperature. In heater cycle mode and odourless air (see \cref{fig:classfication_features}b, left), the sensors exhibited nearly-exponential relationships between the recorded sensor resistances and the hotplate temperatures in the range of \SI{200}{\degreeCelsius} and \SI{400}{\degreeCelsius}, with deviations at lower temperatures (see \cref{fig:supp-setup-methods}c). A small deviation can be observed between the trajectories corresponding to heating and cooling, however this may be attributed to uncertainties in estimating the temperature of the sensor, which is close to but not necessarily equal to that of the hotplate. The resistances returned to their initial values after a completed cycle, without significant hysteresis. 

\subsection*{Odour delivery setup}
\subsubsection*{Reagents}
All odourants were obtained in their pure liquid form from Sigma-Aldrich, and contained in 15-ml glass vials (27160-U, Sigma-Aldrich). The odorants ethyl butyrate, isoamyl acetate and cineol were diluted 1:5 with mineral oil, while 2-heptanone was diluted 1:20 with mineral oil. 2-heptanone was diluted to a lower concentration as a 1:5 dilution saturated some of the gas sensors. 

\subsubsection*{Olfactometer}
Odours were presented using a custom made olfactometer capable of constructing temporally complex stimuli with temporal bandwidths of up to \SI{500}{Hz} (\cref{fig:enose}d). This temporal olfactory delivery device (TODD) has been outlined previously\cite{Ackels2021, Dasgupta4278}. The device consisted of 8 independent channels which contained either odour (diluted with mineral oil) or pure mineral oil. These 8 channels were grouped into two sets of 4. Each set of four consisted of an odour manifold, which contained odours or pure mineral oil in glass vials which were fed by a common air flow. Each channel in this odour manifold was fed into its own high speed valve on a separate valve manifold. Each high speed valve could be opened and closed at frequencies of up to \SI{500}{Hz}. On each valve manifold, one of the channels containing pure mineral oil was set to remain open indefinitely, acting as a 'carrier' valve (grey valves in \cref{fig:enose}d). When a trial was triggered, this carrier valve flow was reduced in accordance with the amount of additional airflow generated by the other valves on the manifold, therefore maintaining a continuous rate of air flow through the system. In some cases, the carrier valve was not simply reduced, but was used to generate temporally complex airflow to compensate for the temporal patterns generated using the other valves in the system. Signals to the valves were convolved with a high frequency \SI{500}{Hz} continuous signal, referred to as 'shattering'. This shattering was included as it has been previously found to improve the temporal fidelity of the resultant odour signal. The airflow to the TODD was maintained at a rate of \SI{1}{\liter \per \min} using a custom closed loop-feedback flow controller.

\subsubsection*{Calibration}
To ensure a continuous total airflow whilst maintaining a high signal fidelity, prior to the electronic nose recording session, the output of the olfactometer was measured with both a PID (200B miniPID, Aurora Scientific) and flowmeter (AWM5101VN, Honeywell). The PID was positioned a short distance away from the output of the olfactometer (> 2cm) and calibration trials were presented. These were selected in a way to making sure to cover all the different valve combinations of the experiment. The PID response to presented odours was measured and the fidelity estimated. 
If the fidelity was found to be too low, the rate of flow into each channel was tuned to increase the fidelity. Next, the PID was replaced with the flowmeter, and the same selected calibration trials were presented. If the rate of flow varied during the trial presentation, the compensatory flow or carrier flow was modulated to return the net flow back to pre-trial levels. The flow through the odour valve was kept constant so as to not alter the odour signal fidelity. Airflow was modulated by altering the duty-cycle of the valve shattering. Once there was no visual change in the rate of flow between the trial and pre-trial levels the flowmeter was removed and the olfactometer was deemed to be calibrated.

\subsubsection*{Fidelity Calculations}
For quantifying the olfactometers' temporal fidelity after calibration, we deployed single-odour pulse trains of different frequencies and obtained simultaneous PID and flow meter recordings. Here, the odourant Ethyl Butyrate was used, as its ionisation energy is well-suited for the used PID. In particular, at $t=\SI{0}{\second}$ the carrier flow valves opened while odour valves remained closed, for a duration of $\SI{10}{\second}$. At $t=\SI{10}{\second}$ the odour valve and odourless compensation valve deployed anti-correlated pulse trains of various frequencies, for $\SI{2}{\second}$ (see \cref{fig:supp-setup-methods}a (left) for the \SI{10}{\hertz} example). The fidelity for each square pulse was calculated as the value of peak to trough, normalised to the peak to the baseline value. The fidelities shown in \cref{fig:supp-setup-methods}a (right) were computed as the mean and standard-deviation across all square pulses fidelities of a particular modulation frequency. 

\subsection*{Experimental protocol}
\subsubsection*{Electronic nose placement}
The electronic nose was attached to a movable arm and fixed in place downstream of the olfactometer outlet, with a distance of approximately \SI{3}{cm} from outlet nozzle to the gas sensors. To ensure that the gas flow reached all the sensors on the board, we fine-tuned the alignment of the electronic nose with respect to the nozzle by trial-and-error until a strong response was obtained on all channels. 

\subsubsection*{Heater modulation and odour delivery protocol}
Three experiments with different sensor heater conditions were performed (see \cref{fig:supp-setup-methods}d for the conditions over time):
\begin{enumerate}
    \item Sensor 1-8: \SI{50}{\milli\second} cycles between \SI{150}{\degreeCelsius} and \SI{400}{\degreeCelsius}. 
    \item Sensor 1-4: constant temperature of \SI{400}{\degreeCelsius}, 
    Sensor 5-8: \SI{50}{\milli\second} cycles between \SI{150}{\degreeCelsius} and \SI{400}{\degreeCelsius}.
    \item Sensor 1-4: constant temperature of \SI{400}{\degreeCelsius},
    Sensor 5-8: \SI{200}{\milli\second} cycles between \SI{150}{\degreeCelsius} and \SI{400}{\degreeCelsius} (not used).
\end{enumerate}
For each heater condition, odour stimuli of different pulse widths and concentrations (controlled by adjusting the shattering duty cycle of odour and mineral oil valves) were presented. After each odour stimulus, there was a \SI{30}{\second} recovery phase before the next stimulus onset. The set of different stimuli included:
\begin{itemize}
    \item 4 odours + two control ('blank') vials, 50 repetitions \SI{1}{\second} odour pulses, at 100\% concentration. 
    \item 4 odours, 20 x \SI{1}{\second} odour pulses, at concentrations of 20\%, 40\%, 60\%, and 80\%. 
    \item 4 odours, 5 repetitions of shorter odour pulses in the range [10, 20, 50, 100, 200, 500] \SI{}{\milli\second}, at 100\% concentration. 
    \item 2x6 odour pairs, 5 repetitions of \SI{1}{\second} anti-correlated pulse trains, at frequencies in the range [1, 2, 5, 10, 20, 40, 60] \SI{}{\hertz}, at 100\% concentration.
    \item 6 odour pairs, 5 repetitions of \SI{1}{\second} correlated pulse trains, at frequencies in the range [1, 2, 5, 10, 20, 40, 60] \SI{}{\hertz}, at 100\% concentration.
\end{itemize}
Within each experimental run, all the stimuli were presented in a fully randomised order. \cref{fig:supp-setup-methods}e shows the distribution of odours over time, binned in \SI{1}{\hour} time intervals. A statistical ${\chi}^2$ test was performed, confirming that the null hypothesis can't be rejected ($p=0.364$), i.e. that the trials are in fact randomised. Importantly, the odour delivery protocol has not been synchronised with the sensor heater modulation phase.

\subsubsection*{PID recordings and odour onset/offset determination}
A shortened version of the odour delivery protocol was deployed and recorded with the PID. 
\cref{fig:supp-setup-methods}f displays a PID response to a \SI{1}{\second} isoamyl acetate pulse. For all the odours, the mean and standard deviation of the pre-stimuls baseline were computed, and a threshold of 4 times the standard deviation ($4 \sigma$) defined. This was used to estimate an upper bound for the time from theoretical stimulus onset to odour exposure at the sensing site, and a lower bound from theoretical stimulus offset to the purging of the sensor site. \cref{fig:supp-setup-methods}g displays all the extracted onset and offset values, indicating that the odour may reach the sensor as rapidly as in \SI{10}{\milli\second}, while the purging may take several hundreds of milliseconds. While PIDs are extremely fast, they too have a finite and odour dependent response time, thus the actual times may be shorter than this. 

\subsection*{Pulse classification analysis}

\subsubsection*{Feature extraction and validation}
For evaluating what data features may be most suitable for the rapid classification of short odour pulses, we used experiment B, where sensor 1-4 were operated at a constant temperature of \SI{400}{\degreeCelsius}, and sensor 5-8 employed heater cycles of \SI{50}{\milli\second} in the range of \SI{150}{\degreeCelsius} and \SI{400}{\degreeCelsius}. 

Different sensor data features were used and evaluated: Data windows of \SI{50}{\milli\second} starting at a given time $t$ relative to the simulus onset at $t_{onset}=\SI{0}{\second}$ were used to extract 
1. raw data from constant heater sensors, 
2. pre-stimulus ($t_{pre}=-\SI{5}{\second}$) baseline subtracted data from constant heater sensors, 
3. raw data from cycled heater sensors, and 
4. pre-stimulus ($t_{pre}=-\SI{5}{\second}$) baseline normalised data from cycled heater sensors. 
For the normalisation in the latter, the procedure is illustrated in \cref{fig:supp-fastclassification-methods}c and described in the following. The extraction of the feature $G(\mathcal{D}_s, t)$ 
can be summarised as applying a chain of a sensor-wise logarithmic transformation and a maximum scaling, to both a \SI{50}{\milli\second} baseline data snippet before stimulus onset (here, $t_{pre}=-\SI{5}{\second}$) and to a snippet at time $t$, and then computing their vector difference:
\begin{equation}
\centering
    G(\mathcal{D}_{s},t) =  
    \frac{\log(H(\mathcal{D}_{s},t))}
    {\max(\log(H(\mathcal{D}_{s},t)))}  -  
    \frac{\log(H(\mathcal{D}_{s}, t_{pre}))}
    {\max(\log(H(\mathcal{D}_{s}, t_{pre})))}
\label{eq:classification_features}
\end{equation}
Here, $H(\mathcal{D}_{s}, t)$ describes a kernel extracting data from the sensor recordings $\mathcal{D}_{s}$, starting at time $t$ and stopping after one full heater cycle (e.g. 50 measurements). The sensor index is denoted as $s$.

The data was split into one set for training \& validation, and one set for testing --- with a ratio of 60\% to 40\% (see \cref{fig:supp-fastclassification-methods}a). The former was used to train and validate a k-NN classifier using different features via cross-validation. The latter was used to evaluate the performance with the different features used. 

Classifier training was performed on data features from sensor responses between \SI{500}{\milli\second} and \SI{1000}{\milli\second} after stimulus onset, where the stimuli were \SI{1000}{\milli\second} odour pulses of the gases 2H, EB, IA, Eu and blank, at 100\% concentration. Testing was performed on equivalent data features, however now the stimuli concentration was sampled in the range [20, 40, 60, 80, 100]\%, and the blank class was omitted. \cref{fig:supp-fastclassification-methods}d displays the achieved performance using the different data features. The normalised cycled-heater data feature $G(\mathcal{D}_{s},t)$ outperforms the other tested features, both in accuracy at 100\% concentration, as well as for the reduced concentrations. For clarity, \cref{fig:classfication_features}e shows a subset of \cref{fig:supp-fastclassification-methods}d.

\subsubsection*{Dynamic pulse classification}
For the dynamic classification of short odour pulses, experiment A was used with all 8 sensors modulated on a \SI{50}{\milli\second} period betweeen \SI{150}{\degreeCelsius} and \SI{400}{\degreeCelsius}. Again, a 60\% vs. 40\% split for training \& validation vs. testing was performed, where the former was used to determine a suitable classifier and its hyper-parameters via cross-validation, while the latter served to evaluate the performance of the classifier (see \cref{fig:supp-fastclassification-methods}b). 
Data features were extracted as described in Eq. \cref{eq:classification_features}. For training, the underlying data for the features are the sensor responses for the subsequent \SI{2000}{\milli\second} after the onset of \SI{1000}{\milli\second} odour stimuli, for concentrations in the range [20, 40, 60, 80, 100]\%. The data features were labelled according to their time $t$ as follows:
\begin{algorithm}[H]
\caption{Training data labelling procedure}
\begin{algorithmic}
\IF{$t_{onset} \leq t < t_{offset} - \tau + d$}
    \STATE Feature is fully within measurable odour pulse.
    \IF{$\mathcal{O}_{stimulus} \text{ is not 'blank'}$}
        \STATE $y = \mathcal{O}_{stimulus}$
    \ELSE
        \STATE $y = \text{'blank'}$
    \ENDIF
\ELSIF{$t_{offset} + d \leq t$}
    \STATE Feature is fully after measurable odour pulse.
    \STATE $y = \text{'blank'}$
\ELSE
    \STATE Feature timing is ambiguous; exclude data feature from training set.
    \STATE $y = \text{'rejected'}$    
\ENDIF
\end{algorithmic}
\end{algorithm}
\noindent 
Here, $t_{onset}=\SI{0}{\milli\second}$ and $t_{offset}=\SI{1000}{\milli\second}$ are the stimulus onset and offset respectively, $\tau=\SI{50}{\milli\second}$ is the feature duration, $d=\SI{10}{\milli\second}$ is the upper bound stimulus delay as computed earlier, $\mathcal{O}_{stimulus}$ is the stimulus odour of the corresponding trial, and $y$ is the prescribed label of the data feature in question. This procedure is illustrated in \cref{fig:fastclassification}a.

For the normalised data features, several classification algorithms were trained and validated via five-fold stratified cross-validation. The best performing algorithm with corresponding hyper-parameters was selected, which here was a Support Vector Machine (SVM) classifier with radial basis function kernel \cite{vert2004primer} ($C=1e3$, $\gamma=1e-4$, balanced class weight). Ultimately, an ensemble classifier was composed from the five SVMs trained on each split. 

For testing how well the trained classifier performs on shorter odour pulses, the features were extracted from sensor response data for the subsequent \SI{2000}{\milli\second} after the onset of odour stimuli of different durations, at 100\% concentration. The stimulus durations fall within the set \{10, 20, 50, 100, 200, 500, 1000\}\SI{}{\milli\second}. For each data feature, the classifier predicted the odour, which is illustrated as a raster plot in \cref{fig:fastclassification}b. The predicted odours $y$ were compared against the actual stimulus odour $\mathcal{O}_{stimulus}$ and divided in predicting 'correct odour', 'wrong odour' and 'no odour', resulting in \cref{fig:fastclassification}c. To extract the accuracy for each pulse duration, a confusion matrix was composed by --- for each trial --- comparing the most predicted non-blank class against $\mathcal{O}_{stimulus}$, across multiple trials. The on- and offset times correspond to the elapsed time from odour onset to first non-blank prediction, and from odour offset to first 'blank' prediction, respectively. 

An analogous procedure was followed for testing the trained classifier on anti-correlated patterns of odour pairs, resulting in predictions over time, as shown in \cref{fig:supp-fastclassification-methods}e \& \cref{fig:supp-fastclassification-methods}f.

\subsection*{Temporal structure analysis}
For the temporal structure analysis, i.e. the determination of the frequency and the phase-shift of the two-odour pulse trains, the constant heater sensor data (i.e. sensors 1-4) of experiment B and C were used. In particular, experiment B was used for training and validation (i.e. finding and evaluating a suitable data feature and classification algorithm), where experiment C was used for testing, see \cref{fig:supp-temporaldiscrimination-methods}a for an illustration. 

\subsubsection*{Feature extraction}
For each data trial, we extracted sensor data $\mathcal{D}_{s}$ from $t=t_{onset}$ to $t=t_{offset}+b$, where $b=\SI{100}{\milli\second}$ to account for the stimulus delay and potential sensor lag. The data was then log-transformed and differentiated before applying a discrete Fourier transformation $\mathcal{F}(.)$, using the fast Fourier Transformation algorithm \cite{cooley1965algorithm}:
\begin{equation}
    L(\mathcal{D}_{s}) = \mathcal{F} \left(\frac{d}{dt} \log(\mathcal{D}_{s}) \right)
\end{equation}
All triplets [frequency, magnitude, phase] were extracted, and sorted according to the magnitude. For each of the four sensors, the triplet with the highest magnitude was selected, collectively composing a 12-dimensional data feature. 
\subsubsection*{Temporal structure classification}
The data features and potential classifiers were evaluated on a 10-fold cross-validation using the training \& validation data (i.e. experiment B) and the three different tasks described earlier. 
We decided on utilising a Random Decision Forest (RDF) classifier \cite{ho1995random} (balanced class weight, $N_{tree}=100$), and on using the same 12-dimensional data feature for all tasks. See \cref{fig:supp-temporaldiscrimination-methods}b-d for an evaluation of the data features. For each cross-validation training split we trained a RDF classifier, then combined them to form an ensemble classifier for each task, which was finally evaluated on the testing data (i.e. experiment C). For the validation using hotplate temperature and PID data, the analogous pipeline was used, except that we omitted the log-transformation. 

\subsection*{Comparing electronic nose performance with mouse performance}
Performance analysis of the electronic nose to discriminate odour correlation structure was carried out in the style of a previously published experimental dataset (See Ackels et al. 2021 \cite{Ackels2021}). This allowed for a direct comparison of the electronic nose performance with that of mice during an operant conditioning task. A complete description of the experimental conditions and data analysis can be found in the original paper. In brief, two cohorts of up to 25 mice were housed in a common home cage system \cite{erskine2019autonomouse} that is used as an automated operant conditioning setup. Mice were trained to discriminate perfectly correlated from perfectly anti-correlated odour stimuli switching at frequencies ranging from $\SI{2}{\hertz}$ to $\SI{81}{\hertz}$. Task frequency was randomized from trial to trial. Odours were presented with a multi-channel high-speed odour delivery device similar to the one used in this manuscript. During a go/no-go task animal performance was rated based on their lick responses to $S+$ (rewarded) and $S-$ (unrewarded) stimuli. For roughly half of all mice, the correlated pattern was $S+$ and the anti-correlated pattern was $S-$ . In the other half of the group this reward valence was reversed. All stimuli were 2 s long. A water reward could be gained by licking so that licking was detected for at least 10\% of the stimulus time during an $S+$ presentation (a ‘Hit’). Licking for the same amount of time during $S-$  presentation resulted in a timeout interval of $\SI{7}{\second}$. In all other response cases, the inter-trial interval was $\SI{3}{\second}$ and no water reward was delivered. All behavioural performance within a specified trial bin was calculated as a weighted average of $S+$ versus $S-$  performance:
\begin{equation}
    \text{Performance} = \frac{(Hit/S+) + (CR/S-)}{2}
\end{equation}
in which $S+$ is the total number of rewarded trials, $S-$ is the total number of unrewarded trials, $Hit$ is the total number of rewarded trials in which a lick response was detected, and $CR$ (correct rejection) is the total number of unrewarded trials in which no lick response was detected.

\bibliographystyle{unsrtnat}

\section*{Acknowledgements}
The authors would like to thank Y. Zhang for assisting with the odourant samples, as well as M. Psarrou, S. Sutton, Y. Bethi, N. Ralph and P. Hurley for fruitful discussions. 

\subsection*{Funding}
Part of this work was funded by an NSF/MRC award under the Next Generation Networks for Neuroscience initiative (NeuroNex Odor to action, NSF \#2014217, MRC \#MR/T046759/1). Part of this work was supported by a Boehringer Ingelheim Fonds PhD fellowship to CDC.

\subsection*{Author contributions statement}
MS, AS, ND, DD, TW, SR and AvS conceptualised the study.
DD designed the electronic nose hardware.
DD and ND developed the electronic nose software.
ND, DD, TW, SR and CDC conducted the investigations and experiments.
ND and DD performed formal analysis.
ND created the visualisations.
MS, AS and AvS acquired funding.
ND drafted the original writing.
Everyone participated in the review and editing process.

\subsection*{Competing interests}
The authors declare no competing interests.

\clearpage

\setcounter{figure}{0}
\renewcommand{\thefigure}{S\arabic{figure}}

\section*{Supplementary Materials}

\begin{figure}[h!]
    \centering
    \includegraphics[width=\linewidth]{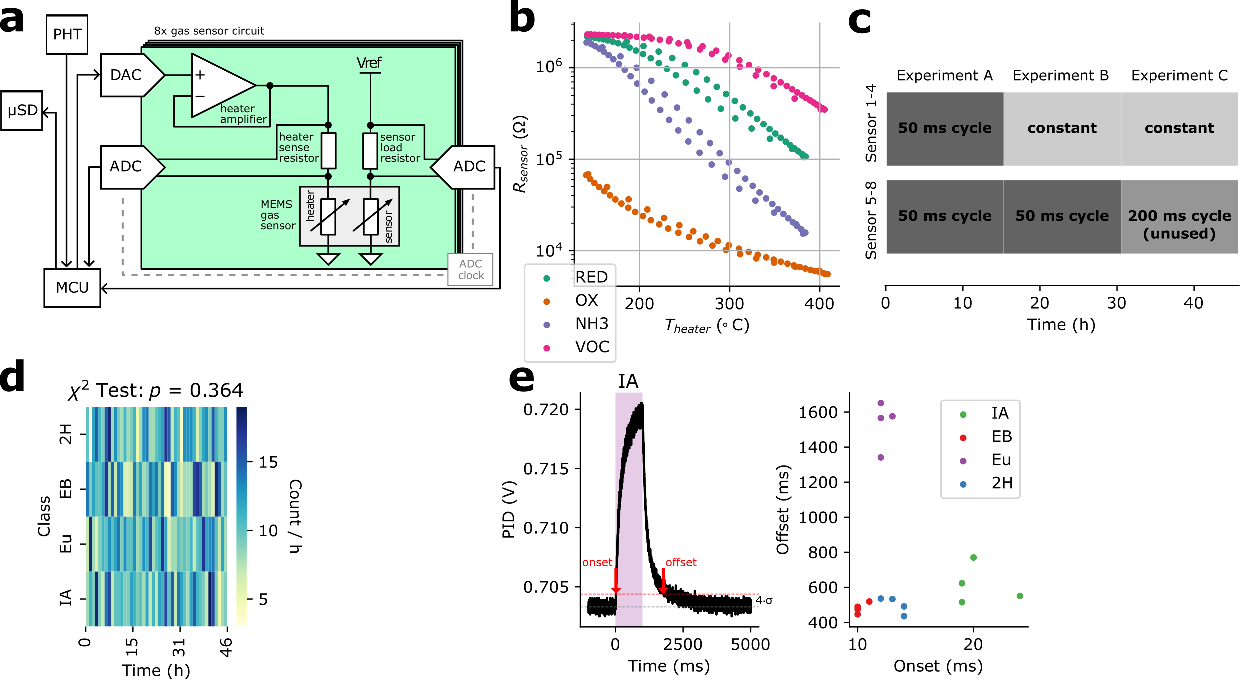}
    \caption{\textbf{Supplementary figure for experimental setup.}
    \textbf{a}, Electronic nose design, displaying how the microcontroller unit (MCU) sets and reads out the sensor heaters in a closed loop, while reading out the analyte dependent sensor resistances. Further, the MCU connects to an environmental sensor (PHT) and a micro SD card. 
    \textbf{b,} R-T curve of a \SI{50}{\milli\second} temperature cycle between \SI{150}{\degreeCelsius} and \SI{400}{\degreeCelsius} without external stimulus, displaying how the sensor response closely follows the hotplate temperature. 
    \textbf{c,} Different sensor hotplate settings over time. For each experiments, all the stimuli were presented in randomised order.
    \textbf{d,} Heatmap depicting the distribution of odour presentations over a set of 1 hour time intervals. A ${\chi}^2$ test was performed to assess the randomness of class distribution over time intervals, with the computed p-value indicated as '$p$'.
    \textbf{e,} PID response to a \SI{1}{\second} isoamyl acetate pulse. 
    Grey-dotted and red dotted lines denote mean of pre-stimulus baseline and 4 standard deviations threshold respectively. Where the response crosses the threshold upwards (downwards), the odour onset (offset) is registered. 
    \textbf{f,} Extracted odour onsets (w.r.t. $t=\SI{0}{\milli\second}$) and offsets (w.r.t. $t=\SI{1000}{\milli\second}$) for \SI{1000}{\milli\second} pulses of different odours.
    For all experiments, the odourants are abbreviated as follows. IA: isoamyl acetate; EB: ethyl butyrate; Eu: cineol; 2H: 2-heptanone; blank: odourless control.
    }
    \label{fig:supp-setup-methods}
\end{figure}

\begin{figure}
    \centering
    \includegraphics[width=\linewidth]{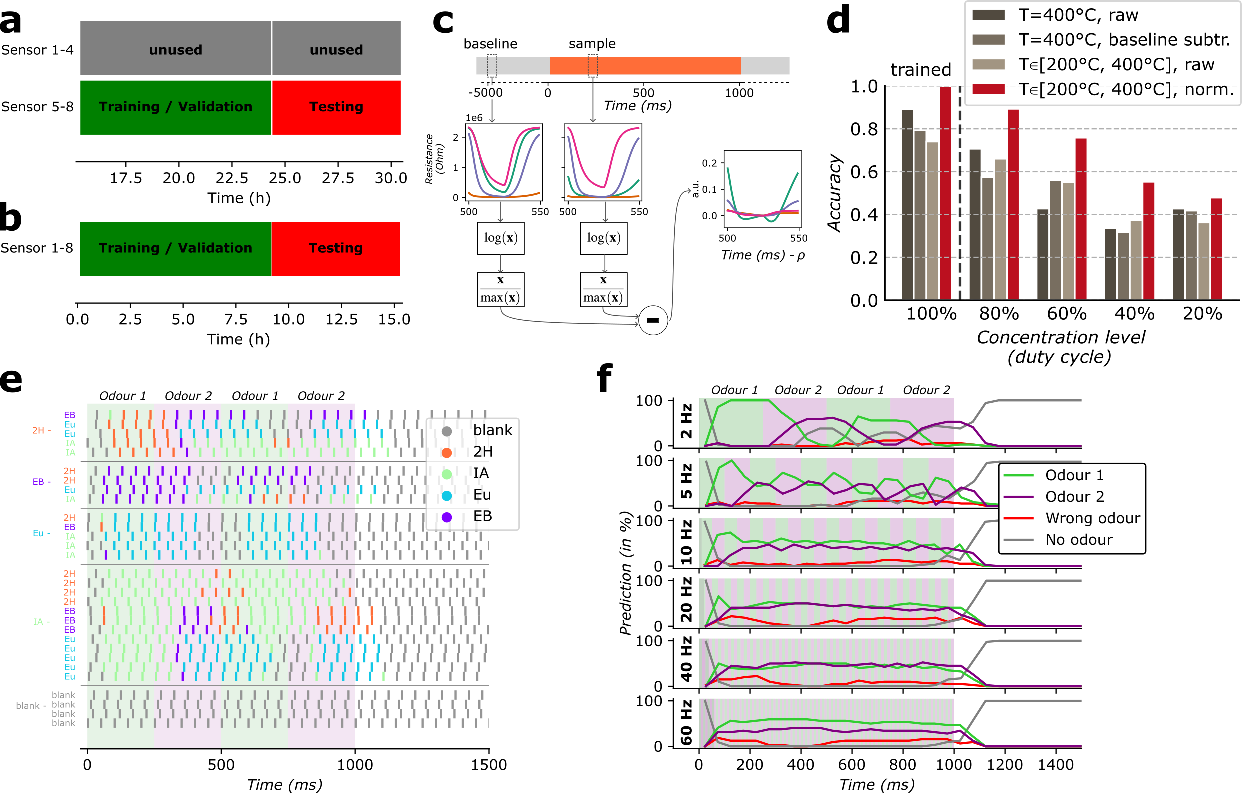}
    \caption{\textbf{Supplementary figure for fast odour classification.}
    \textbf{a,} Data splitting for robustness analysis of the rapid heater modulation data features (see \cref{fig:classfication_features}e). 
    \textbf{b,} Data splitting for evaluating the dynamic classification of millisecond odour pulses (see \cref{fig:fastclassification}). 
    \textbf{c,} Normalisation procedure for the heater modulation data feature. Time shifted by cycle phase $\rho$ w.r.t. odour onset, for visual guidance only. 
    \textbf{d,} Accuracy scores for a k-nearest neighbours (k-NN) classifier trained on \SI{50}{\milli\second} data features from \SI{1000}{\milli\second} odour pulses at full concentration, and tested on \SI{50}{\milli\second} features from \SI{1000}{\milli\second} odour pulses at different concentration levels (tuned by adjusting the duty cycle of the micro-valves). Features are compared for constant heater sensor readings (raw and baseline-normalised) and cycled heater sensor readings (raw and normalised, as described in c)) 
    \textbf{e,} Odour stimulus classification for anti-correlated odour patterns over time. An RBF-kernel SVM classifier was trained on \SI{50}{\milli\second} features from \SI{1000}{\milli\second} odour pulses, and tested on odour anti-correlated odour patterns of various frequencies. Shown here is is a \SI{2}{\hertz} pattern. 
    \textbf{f,} Classification correctness over time (evaluated via the true odour presence), for anti-correlated odour patterns of different switching frequencies. 
    }
    \label{fig:supp-fastclassification-methods}
\end{figure}

\begin{figure}
    \centering
    \includegraphics[width=\linewidth]{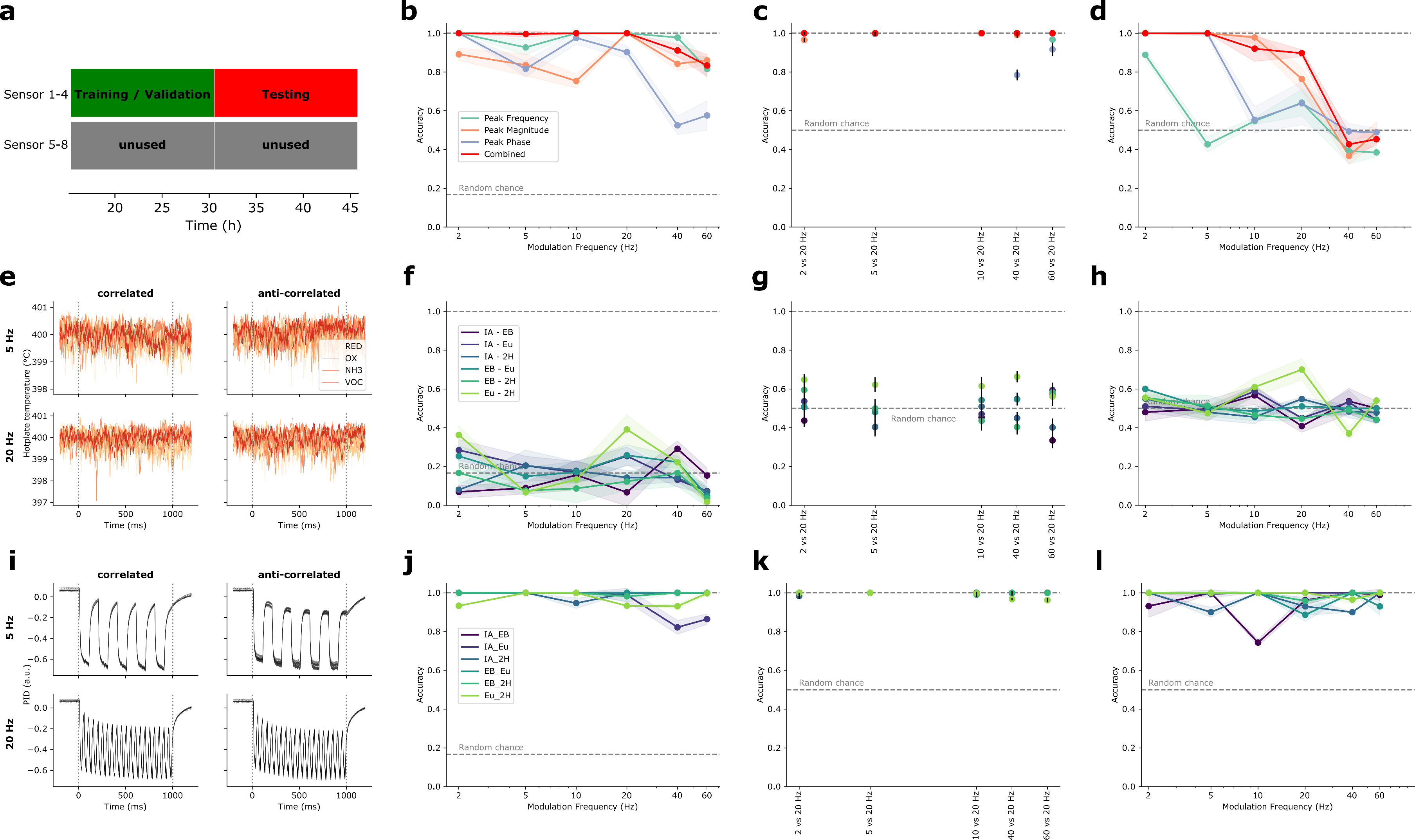}
    \caption{
    \textbf{Supplementary figure for temporal pattern discrimination.}
    \textbf{a,} Data splitting for evaluating the temporal pattern discrimination performance (see \cref{fig:temporal_combined}). 
    \textbf{b-d,} Validation accuracy plots for different extracted DFT-spectrogram peak features using the MOx gas sensor resistances. \textbf{b,} Modulation frequency classification, 
    \textbf{c,} pairwise modulation frequency classification, and 
    \textbf{d,} correlated vs anti-correlated modulation discrimination. 
    \textbf{e,} MOx heater temperature values for different odour modulations. Here shown are data for the odour pair IA (isoamyl acetate)- EB (ethyl butyrate), 5 trials each for 5 Hz correlated, 5 Hz anti-correlated, 20 Hz correlated and 20 Hz anti-correlated respectively.
    \textbf{f-h,} Test accuracy plots for different odour pair modulations, using the MOx heater temperature values. 
    \textbf{i,} Photoionisation Detector (PID) responses for different odour modulations (odour stimuli as in \textbf{e}). 
    \textbf{j-l,}, Test accuracy plots for different odour pair modulations, using the Photoionisation Detector (PID) responses. 
    For all the classification tasks, an ensemble of Random Forest Classifiers was used. The mean and error estimations arise from repeating training and testing with different random seeds. 
    }
    \label{fig:supp-temporaldiscrimination-methods}
\end{figure}

\end{document}